\documentclass[twocolumn,preprintnumbers,amsmath,amssymb]{revtex4}
\usepackage{graphicx}
\usepackage{dcolumn}
\usepackage{epsfig}
\usepackage{bm}
\def\be{\begin{equation}}
\def\ee{\end{equation}}
\def\bea{\begin{eqnarray}}
\def\eea{\end{eqnarray}}
\def\nnb{\nonumber}

\begin{document}


\title{Transition Radiation by Neutrinos at an Edge of Magnetic Field}

\author{A. Ioannisian$^{a,b}$}
\author{N. Kazarian$^{b}$}
\affiliation{ $^a$ Yerevan Physics Institute, Alikhanian Brothers
 2, Yerevan-36,
Armenia\\
$^b$ Institute for Theoretical Physics and Modeling, Yerevan-36,
Armenia
 }

\begin{abstract}
We calculate the transition radiation process $\nu \to \nu \gamma$
at an interface of magnetic field and vacuum. The neutrinos are taken to be with
only standard-model couplings. The magnetic field fulfills the dual
purpose of inducing an effective neutrino-photon vertex and of
modifying the photon dispersion relation. The transition radiation
occurs when at least one of those quantities  have different
values in different media. The neutrino mass is ignored due to its
negligible contribution. We  present a result for the probability
of the transition radiation which is both accurate and analytic.
\\
\end{abstract}


\maketitle

{\it \bf Introduction.-} 

In many astrophysical environments the absorption, emission, or
scattering of neutrinos occurs in media, in the presence of 
magnetic fields \cite{Raffelt:1996wa} or at the interface 
of two media. Of particular conceptual interest
are those reactions which have no counterpart in vacuum, notably
the plasmon decay $\gamma\to\bar\nu\nu$, the Cherenkov and
transition radiation processes $\nu\to\nu\gamma$. These reactions
do not occur in vacuum because they are kinematically forbidden
and because neutrinos do not couple to photons. In the presence of
a media (or a magnetic field), neutrinos acquire an effective
coupling to photons by virtue of intermediate charged particles.
In uniform media (or external field) the dispersion relations are
modified of all particles so that phase space is opened for
neutrino-photon reactions of the type $1\to 2+3$. The violation of
the translational invariance at the direction from one media into
another leads to the non conservation of thr  momentum at the same
direction so that transition radiation becomes kinematically
allowed.

The theory of the transition radiation by charged particle  has been diveloped in \cite{Ginzburg:1945zz}\cite{Garibyan:1959}.
It those articles authors used classical theory of electrodynamics. In \cite{Garibyan:1960} 
the quantum field theory was used for describing the phenomenon. 
The neutrinos have very tiny masses. 
Therefore one has to use the quantum field theory approach in order to study transition radiation by neutrinos.

The plasma process $\gamma\to\bar\nu\nu$ was first studied in \cite{Adams:1963zz}. The
\hbox{$\nu$-$\gamma$}-coupling is enabled by the presence of the
electrons of the background medium, and the process is
kinematically allowed because the photons acquire essentially an
effective mass. The plasma process is the dominant source for
neutrinos in many types of stars and thus is of great practical
importance in astrophysics~\cite{Raffelt:1996wa}. 
In a plasma there are electromagnetic excitations, namely, the longitudinal plasmons, $\gamma_L $, which four-momentum are space like for certain energies. The Cherenkov decay $\nu \to \nu \gamma_L$ was studied in \cite{Oraevsky:1987cu}.

The presence of a magnetic field induces an effective
$\nu$-$\gamma$-coupling. The Cherenkov decay in a magnetic field
was calculated in  \cite{Ioannisian:1996pn}. The $\gamma\to\bar\nu
\nu$ decay rate was calculated in \cite{DeRaad}, assuming that phase space is
opened by a suitable medium- or field-induced modification of the
photon refractive index.

At the interface of two media with different refractive indices 
the transition radiation by standard model neutrinos, $\nu\to\nu\gamma$, was studied in \cite{Ioannisian:2011mf}.
Transition radiation by neutrinos with large magnetic/electric  dipole moment was studied in
\cite{Sakuda:1994zq}.

The magnetic field causes an effective $\nu$-$\gamma$-vertex in the 
standard-model. Also the magnetic field  changes photon's dispersion relation. 
We neglect neutrino masses and medium-induced modifications of their  
dispersion relation due to their negligible role.
Therefore, we study the transition radiation entirely
within the particle-physics standard model.

\begin{figure}[b]
  \includegraphics[height=40mm]{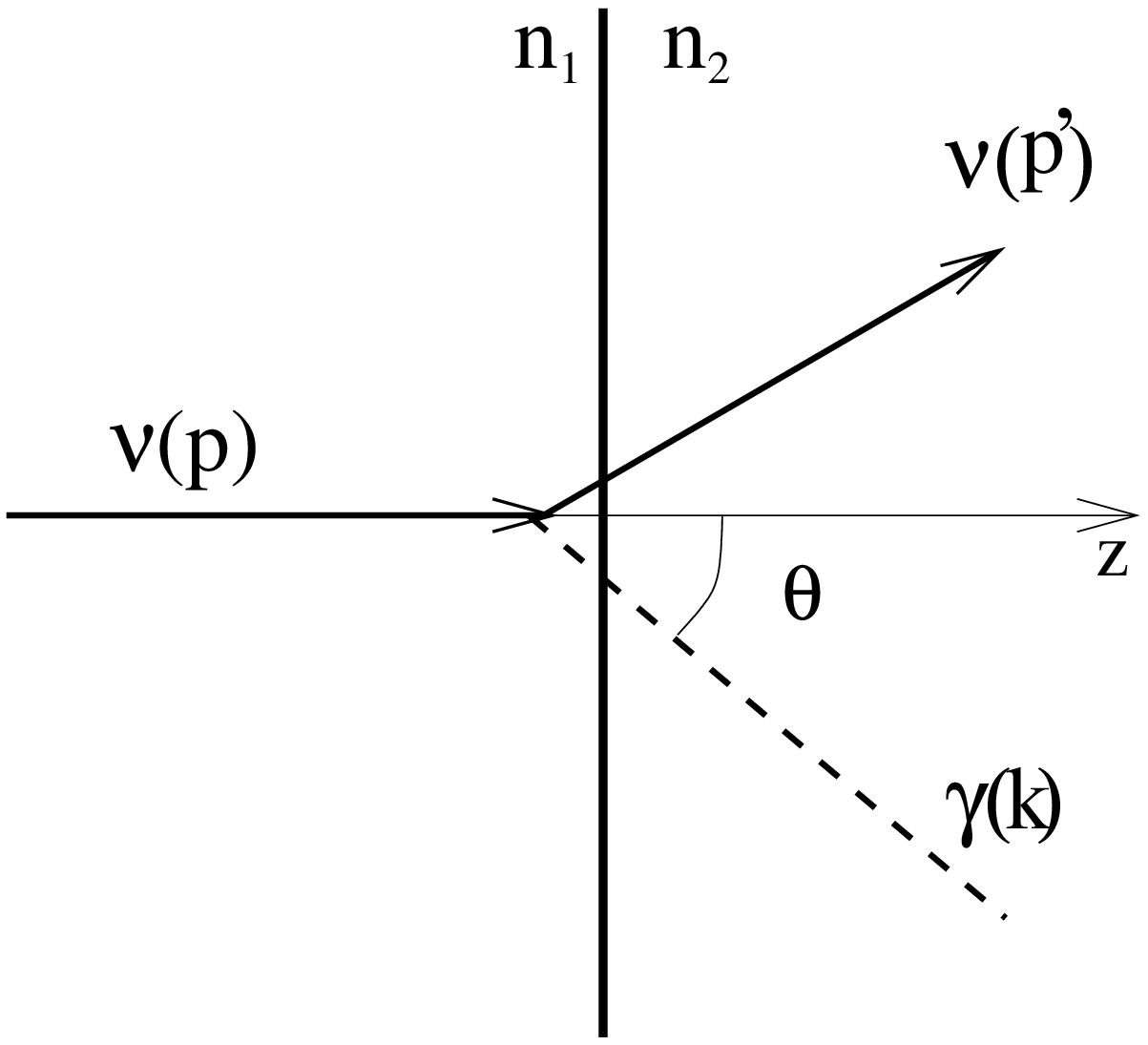}
\caption[...]{Transition radiation by neutrino at an interface of two media with 
refractive indexes $n_1$ and $n_2$.
\label{Fig0}}
\end{figure}

We proceed by deriving a general expression for the
transition radiation rate (assuming a general
$\nu$-$\gamma$-vertex) in quantum field theory. 
We derive the standard-model
effective vertex in a magnetic field, then we
calculate the transition radiation rate by performing 
semi-analytical integrations and summarize our findings.

{\it \bf Transition Radiation.-}

Let us consider a neutrino crossing the interface of two
media with refraction indexes $n_1$ and $n_2$ (see Fig. 1). In
terms of the matrix element ${\cal M}$ the transition radiation
probability  of the process $\nu \to \nu \gamma$ is \cite{Ioannisian:2011mf}
 \bea
  W &=&  {1 \over (2\pi)^3}{1 \over 2E \beta_z}
  \int
 {d^3 {\textbf{p}^\prime} \over 2 E^\prime}{d^3
 {\textbf{k}} \over  2 \omega}
 \sum_{pols}\left|\int_{-\infty}^\infty \! \! \! \! \! \! \! dz
 e^{i(p_z-p_z^\prime-k_z)z}{\cal M}\right|^2  \nnb \\
 && \hspace{1cm} \times \
 \delta(E-E^\prime-\omega)\delta(p_x^\prime+k_x)\delta(p_y^\prime+k_y) \ .
\label{1} \eea Here, $p=(E,{\bf p})$, $p'=(E',{\bf p}')$, and
$k=(\omega,{\bf k})$ are the four momenta of the incoming
neutrino, outgoing neutrino, and photon, respectively and
$\beta_z={p_z / E}$. The sum is over the photon polarizations.

\begin{figure}[b]
  \includegraphics[height=40mm]{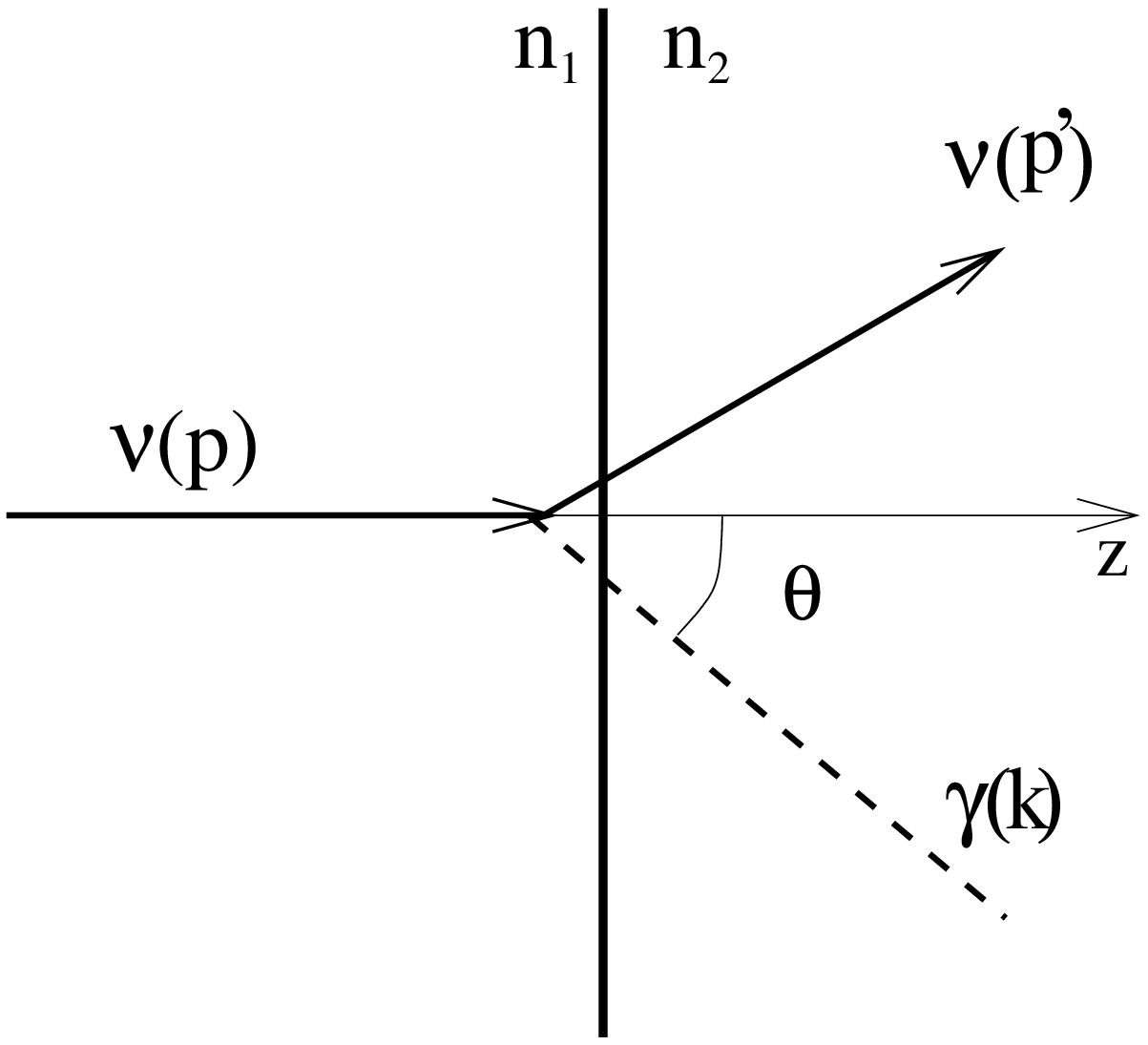}
\caption[...]{Transition radiation by neutrino at the interface of
two media with refractive indexes $n_1$ and $n_2$. \label{Fig0}}
\end{figure}

We shall neglect the neutrino masses and the deformation of its dispersion relations due to the forward scattering.
Thus we assume that the neutrino dispersion relation is precisely
light-like so that $p^2=0$ and $E=|{\bf p}|$.

The transition radiation formation zone length of the medium is
\be
  |p_z-p_z^\prime-k_z|^{-1}.
\ee The integral over $z$ in Eq. (\ref{1}) oscillates beyond the
length of the formation zone. Therefore,  the contributions to the
process from the depths over the formation zone length may be
neglected. The neutrino transfers the z momentum
$(p_z-p_z^\prime-k_z)$ to the medium. Since photon propagation in
the medium suffers from the attenuation(absorption) the formation
zone length must be limited by the attenuation length of the
photons in the medium when the latter is shorter than the
formation zone length.

After integration of (\ref{1})
over $  \textbf p'$ and $z$ we find
 \bea
\label{w1}
W &=&  {1 \over (2\pi)^3}{1\over 8E \beta_z}
  \int
 { |{\textbf{k}}|^2 d |{\textbf{k}}| \over \omega
 E^\prime \beta_z^\prime} \ \sin \theta \ d \theta \ d \varphi \nnb \\
 && \times \! \sum_{pols} \left|{{\cal M}^{(1)} \over
 p_z-p_z^{\prime(1)}-k^{(1)}_z} - {{\cal M}^{(2)} \over
 p_z-p_z^{\prime(2)}-k^{(2)}_z}\right|^2 \! \! ,
 \eea
where $\beta^\prime_z=p'_z / E^\prime$, $\theta$ is the angle
between the emitted photon and incoming neutrino. ${\cal
M}^{(1,2)}$  are matrix elements of the $\nu \to \nu \gamma$ in
each medium. $k^{(i)}_z$ and $p_z^{\prime(i)}$ are $z$ components
of momenta of the photon  and of the outgoing neutrino in each
medium.

In a magnetic field the photon refractive index is not isotropic, and
it depends on the photon polarization. According to Adler's classic
paper~\cite{Adler} there are two eigenmodes of photon propagation, one
with the polarization vector parallel ($\parallel$) and one
perpendicular ($\perp$) to the plane containing ${\bf k}$ and 
${\bf B}$.\footnote{Our definition of $\parallel$ and $\perp$ is 
opposite 
to Adler's~\cite{Adler} who used the photon's magnetic-field vector
to define the polarization.} The refractive
index to lowest order in $\alpha$ in magnetic field is 
$
n_{\parallel,\perp}=1+
\frac{\alpha}{4\pi}\eta_{\parallel,\perp}\sin^2\beta
$, 
where $\beta$ is the angle between ${\bf k}$ and ${\bf B}$.  The
numerical coefficients $\eta_{\parallel,\perp}$ depend on $B$,
$\omega$, and $\beta$ .
For $B={\cal O}(B_{\rm crit})$ they are of order
unity. 
In the matter ( without magnetic field) the index of refraction of the transverse photons is $n=1-{\omega^2 \over 2 \omega_{plasmon}^2}$ and thus $1-n \ll 1$ when $\omega \gg \omega_{plasmon}$.
Therefore, for all situations of practical interest we have
$|n_{\parallel,\perp}-1|\ll 1$.

As it is shown below the main contribution to the process comes from the large formation zone lengths and, thus,
 small angle $\theta$. Therefore, the rate of the process does not depend on the angle between the momenta of the
 incoming neutrino and the boundary surface of two media (if that angle is not close to zero). The integration
 over $\varphi$ drops out and we may replace $d\varphi \to 2\pi$. $k^{(i)}_z$ and $p_z^{\prime(i)}$ have the forms
 \be
 k_z^{(i)}=n^{(i)}\omega \cos \theta , \ \
p_z^{\prime(i)}=\sqrt{(E-\omega)^2-{n^{(i)}}^2\omega^2 \sin^2\theta} \   ,
 \ee
here we used  $n^{(1,2)}={|\textbf k|^{(1,2)}}/\omega$.

Moreover the outgoing photon
propagates parallel to the original neutrino direction.

We do not need to distinguish
between $\omega$ and $|{\bf k}|=n\omega=
\hbox{$\omega[1+{\cal O}(\alpha)]$}$. 
Therefore, to lowest order in $\alpha$ the transition radiation probability
Eq.~(\ref{w1}) is found to be
 \bea
W &=&  {1 \over (2\pi)^2}{1\over 8 E^2}
  \int
 {\omega \  \sin \theta \  d \omega  \  d \theta\over 
 (1-{\omega \over E})}   \nnb \\
 && \times \! \sum_{pols} \left|{{\cal M}^{(1)} \over
 p_z-p_z^{\prime(1)}-k^{(1)}_z} - {{\cal M}^{(2)} \over
 p_z-p_z^{\prime(2)}-k^{(2)}_z}\right|^2 \! \! .
 \eea

{\it \bf The Neutrino-Photon-Vertex}

In a magnetic field, photons couple to neutrinos by the amplitudes
shown in Figs.~2(a) and (b). The electron propagator, represented by a
double line, is modified by the field to allow for a nonvanishing
coupling. It has been speculated that superstrong magnetic fields may
exist in the early universe, but we limit our discussion to field
strengths not very much larger than $B_{\rm crit}=m_e^2/e$ which is
the range thought to occur in pulsars.  Therefore, while in principle
similar graphs exist for $\mu$ and $\tau$ leptons, we may neglect
their contribution. For the same reason we may ignore field-induced
modifications of the gauge-boson propagators.  Moreover, we are
interested in neutrino energies very much smaller than the $W$- and
$Z$-boson masses, allowing us to use the limit of infinitely heavy
gauge bosons.

The $\nu$-$\gamma$-vertex in a magnetic field has been investigated in \cite{Ioannisian:1996pn}. According to that result the matrix element for the $\nu$-$\gamma$ vertex can be written in the form 
\begin{equation}
\label{m}
{\cal M}=-\frac{G_F}{\sqrt{2}\,e}Z\varepsilon_{\mu}
\bar{\nu}\gamma_{\nu}(1-\gamma_5)\nu\,
(g_V\Pi^{\mu \nu}-g_A\Pi_5^{\mu \nu})
\end{equation}
here 
$g_V=2\sin^2\theta_W+\frac{1}{2}$ and 
$g_A=\frac{1}{2}$ for $\nu_e$, and
$g_V=2\sin^2\theta_W-\frac{1}{2}$ and
$g_A=-\frac{1}{2}$ for $\nu_{\mu,\tau}$. 
$\Pi$ is the photon polarization tensor or vector-vector (VV)
response function in the magnetic field, while $\Pi_5$ is the
vector-axial vector (VA) response function.
$\varepsilon$ is the photon
polarization vector and $Z$ its wave-function renormalization
factor. For the physical circumstances of interest to us, the photon
refractive index will be very close to unity so that we will be able
to use the vacuum approximation $Z=1$.

The VA response function is \cite{Ioannisian:1996pn} 
\begin{eqnarray}
\label{p5}
\Pi_5^{\mu \nu}(k)&=&\frac{e^3}{(4\pi)^2m_e^2}
\Bigl\{-C_\|\,k_{\|}^{\nu}(\widetilde{F} k)^{\mu}\nonumber\\
&+&C_\bot\,\Bigl[k_{\bot}^{\nu}(k\widetilde{F})^{\mu}
+k_{\bot}^{\mu}(k\widetilde{F})^{\nu}-
k_{\bot}^2\widetilde{F}^{\mu \nu}\Bigr]\Bigr\},
\end{eqnarray} 
where 
\begin{eqnarray}
\label{c}
C_\|&=&im_e^2\int_0^{\infty}ds\int_{-1}^{+1}
dv\,e^{-is\phi_0}(1-v^2)
\nonumber\\
C_\bot&=&im_e^2\int_0^{\infty}ds\int_{-1}^{+1}
dv\,e^{-is\phi_0}R
\end{eqnarray}
are dimensionless coefficients which are real for $\omega<2m_e$, i.e.\
below the pair-production threshold. In Eq. (\ref{c})
\bea
\phi_0&=&m_e^2+\frac{1-v^2}{4}\,k_{\|}^2+
\frac{\cos zv -\cos z}{2z \sin z}\,k_{\bot}^2, \\
R&=&\frac{1-v \sin zv \sin z - \cos z \cos zv}{\sin ^2 z}, 
\eea
$z=eBs$. $\widetilde{F}^{\mu \nu}=
\frac{1}{2}\epsilon^{\mu \nu \rho \sigma}F_{\rho \sigma}$
with $\epsilon^{0123}=1$ is the dual of the field-strength tensor.

\begin{figure}
\centering\leavevmode
\vbox{
\unitlength=0.8mm
\begin{picture}(60,25)
\put(8,15){\line(-1,1){8}}
\put(8,15){\line(-1,-1){8}}
\put(0,7){\vector(1,1){4}}
\put(8,15){\vector(-1,1){6}}
\multiput(9.5,15)(6,0){3}{\oval(3,3)[t]}
\multiput(12.5,15)(6,0){3}{\oval(3,3)[b]}
\put(31,15){\circle{10}}
\put(31,15){\circle{9}}
\multiput(37.5,15)(6,0){3}{\oval(3,3)[t]}
\multiput(40.5,15)(6,0){3}{\oval(3,3)[b]}
\put(0,10){\shortstack{{}$\nu$}}
\put(18,18){\shortstack{{$Z$}}}
\put(43,18){\shortstack{{$\gamma$}}}
\put(30,11){\shortstack{{e}}}
\put(57,13){\shortstack{{(a)}}}
\end{picture}

\unitlength=0.8mm
\begin{picture}(60,32)
\put(16,15){\line(-1,1){7.5}}
\put(16,15){\line(-1,-1){7.5}}
\put(15,15){\line(-1,1){7}}
\put(15,15){\line(-1,-1){7}}
\put(16,15){\line(-1,1){16}}
\put(16,15){\line(-1,-1){16}}
\put(1,0){\vector(1,1){6}}
\put(16,15){\vector(-1,1){10}}
\multiput(17.5,15)(6,0){3}{\oval(3,3)[t]}
\multiput(20.5,15)(6,0){3}{\oval(3,3)[b]}
\multiput(8.2,8.7)(0,6){3}{\oval(3,3)[l]}
\multiput(8.2,11.7)(0,6){2}{\oval(3,3)[r]}
\put(2,15){\shortstack{{}$W$}}
\put(0,3){\shortstack{{}$\nu$}}
\put(23,18){\shortstack{{$\gamma$}}}
\put(57,13){\shortstack{{(b)}}}
\end{picture}

\vskip3ex

\unitlength=0.8mm
\begin{picture}(60,25)
\put(8,15){\line(-1,1){8}}
\put(8,15){\line(-1,-1){8}}
\put(0,7){\vector(1,1){4}}
\put(8,15){\vector(-1,1){6}}
\put(13,15){\circle{10}}
\put(13,15){\circle{9}}
\multiput(19.5,15)(6,0){3}{\oval(3,3)[t]}
\multiput(22.5,15)(6,0){3}{\oval(3,3)[b]}
\put(0,10){\shortstack{{}$\nu$}}
\put(26,18){\shortstack{{$\gamma$}}}
\put(12,11){\shortstack{{e}}}
\put(57,13){\shortstack{{(c)}}}
\end{picture}
}
\smallskip
\caption[...]{Neutrino-photon coupling in an external magnetic field.
The double line represents the electron propagator in the presence of
a $B$-field. 
(a)~$Z$-$A$-mixing. (b)~Penguin diagram (only for $\nu_e$).
(c)~Effective coupling in the limit of infinite gauge-boson masses.
\label{FigI}}
\end{figure}
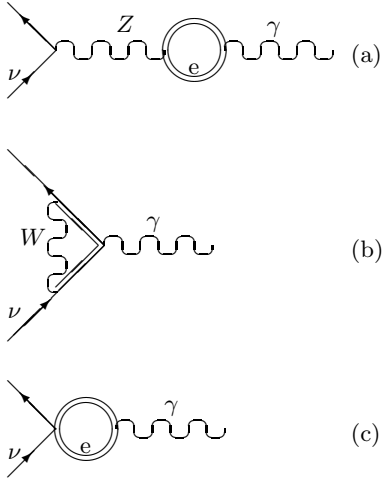

In the presence of matter the $\nu$-$\gamma$-vertex is the sum of the vacuum one (as discussed above) and the matter contribution. It is well known that the last one is proportional to the small parameter $(n-1)^2 \ll 1$. We'll ignore matter contribution to the $\nu$-$\gamma$-vertex  in the present work.


{\it \bf Transition Radiation Probability}

Armed with these results we may now turn to an evaluation of the rate
for $\nu\to\nu\gamma$ at an interface of magnetic field and vacuum. 
It is easy to see that for both photon
eigenmodes the parity-conserving part of the effective vertex
($\Pi^{\mu \nu}$) is proportional to the small parameter
$(n_{\|,\bot}-1)^2 \approx
(\alpha/2\pi)\eta_{\|,\bot}\sin^2\beta$.  It is important to
note that the parity-violating part ($\Pi_5^{\mu \nu}$) is {\it not\/}
proportional to this small parameter for the $\parallel$ photon mode,
while it is proportional to it for the $\perp$ mode.

For neutrinos which propagate perpendicular to the magnetic
field the 
transition radiation probability of $\parallel$ photons is
\begin{equation}
\label{w3}
W=\frac{\alpha G_F^2}{9 \ 2^6 \pi^5}
\left(\frac{B}{B_{\rm crit}}\right)^2 
\int h(B) {\omega^5 \sin \theta d\omega d \theta
\over
( p_z-p_z^{\prime}-k_z)^2}.
\end{equation}
where 
\begin{equation}
h(B)\equiv\frac{9}{16}(C_\|-2C_\bot)^2.
\end{equation}
It turns out that in the range
$0<\omega< 2m_e$ the expression $C_\|/2-C_\perp$ depends only
weakly on $\omega$ so that it is well approximated by its value at
$\omega=0$.  Explicitly, this is found to be
 \begin{equation}
h(B)= 
 \begin{cases}
    (4/25)\,(B/B_{\rm crit})^4 & \text{for $B\ll B_{\rm crit}$}.\\
    1 & \text{for $B\gg B_{\rm crit}$}.
 \end{cases}
 \end{equation}

$p_z^\prime$ in Eq. ~(\ref{w3}) is
\be
p_z^\prime=\sqrt{(E-\omega)^2-n^2\omega^2\sin^2\theta} \ .
 \ee
The maximal allowed angle, $\theta_{max}$ , for the photon
emission is $\pi/2$ when $\omega<{E \over 2}$ and $\sin
\theta_{max} ={E-\omega \over \omega}$ when $\omega>{E \over 2}$.

Now we expand the integrand (\ref{w3}) into the series in small angle, since only in
that case the denominator is small (and the formation zone length
is large). Thus we write the transition probability in the form

\begin{equation}
\label{w4}
W \simeq \frac{\alpha G_F^2}{9 \ 2^6 \pi^5}
\left(\frac{B}{B_{\rm crit}}\right)^2 \! \! \! h(B) \! \! 
\int  \! \! {4 (1-{\omega \over E})^2  \omega^3  \theta d\omega d \theta
\over
( \theta^2+(1-n^2)(1-{\omega \over E}))^2}.
\end{equation}

When  $n < 1$ eq.~(\ref{w4}) can be written in the form
 \be
\label{w5}
W \simeq \frac{\alpha G_F^2}{9 \ 2^6 \pi^5}
\left(\frac{B}{B_{\rm crit}}\right)^2 \! \! \! h(B) \! \! 
\int  \! \! {2 (1-{\omega \over E})  \omega^3   d\omega 
\over
(1-n^2)}.
 \ee

Eqs. (\ref{w4}) and (\ref{w5}) present the main results of the present work.

\vspace{0.3cm}


The index of refraction of photon is greater than 1,  $n>1$, for photon energies  $\omega < 2 \ m_e$  and in pure magnetic field . The pole appears in the denominator under the integral of eq. (\ref{w4}). Therefore outgoing radiation will be a combination of Cherenkov \cite{Ioannisian:1996pn} and transition  (\ref{w4}) ones.  When $0<(1-n) \ll 1$ and we limited the thickness of  magnetic field to d, the resulting radiation probability will be one from eq (\ref{w4} with replacement  of 
$1\over 1-n$ under the integral to $2 \pi \omega$. This probability will exactly coincide with the probability of the cherencov radiation , $d \ \Gamma_{Cher} $, where $\Gamma_{Cher} $ is cherencov radiation rate from \cite{Ioannisian:1996pn}. Thus, in that case both results gave the same answer.

When the process is taking place in presence of some matter (plasma, airÉ), matter contribution to the index of refraction may overcome the magnetic field contribution, and the index of refraction will be less than 1, $n<1$. In that case, only transition radiation will be responsible for outgoing photon according to eq. (\ref{w5}).

Our result is very similar to the electron transition radiation. Our previous work on the subject, transition radiation by neutrino at an interface of matter and vacua  \cite{Ioannisian:2011mf},  has an additional  suppression by the square of the small angle,  $\theta^2$.
 
When the magnetic field has an opposite direction on  each side from the interface,  both sides  give the same sign contribution to the transition radiation amplitude (unlike usual electron transition radiation when they subtract from each other and exactly cancel each other when $n_1=n_2$)  and the resulting probability will be 4 times larger than (\ref{w5}).



\end{document}